\documentclass[pre,aps,superscriptaddress,showpacs,preprint,nofootinbib,floatfix]{revtex4-1}

\usepackage{amssymb,amsmath}
\usepackage{graphicx}
\usepackage{nicefrac}
\usepackage{color}
\usepackage{textcomp,gensymb}
\usepackage{enumerate} 
\usepackage{upgreek}
\usepackage[symbol]{footmisc}

\begin{document}

\title
{Radiation drive temperature measurements in aluminium via radiation-driven shock waves: Modeling using self-similar solutions}
\author
{Shay I. Heizler}
\email{highzlers@walla.co.il}
\affiliation{Department of Physics, Nuclear Research Center-Negev, P.O. Box 9001, Beer-Sheva 8419001, Israel}
\author
{Tomer Shussman}
\affiliation{Department of Plasma Physics, Soreq Nuclear Research Center, Yavne 8180000, Israel}
\author
{Moshe Fraenkel}
\affiliation{Department of Plasma Physics, Soreq Nuclear Research Center, Yavne 8180000, Israel}

\begin{abstract}

We study the phenomena of radiative-driven shock waves using a semi-analytic model based on self similar solutions of the radiative hydrodynamic problem. The relation between the hohlraum drive temperature $T_{\mathrm{Rad}}$ and the resulting ablative shock $D_S$ is a well-known method for the estimation of the drive temperature. However, the various studies yield different scaling relations between $T_{\mathrm{Rad}}$ and $D_S$, based on different simulations. In [T. Shussman and S.I. Heizler, Phys. Plas., 22, 082109 (2015)] we have derived full analytic solutions for the subsonic heat wave, that include both the ablation and the shock wave regions. Using this self-similar approach we derive here the $T_{\mathrm{Rad}}(D_S)$ relation for aluminium, using the detailed Hugoniot relations and including transport effects. By our semi-analytic model, we find a spread of $\approx 40$eV in the $T_{\mathrm{Rad}}(D_S)$ curve, as a function of the temperature profile's duration and its temporal profile. Our model agrees with the various experiments and the simulations data, explaining the difference between the various scaling relations that appear in the literature.

\end{abstract}

\maketitle

\section{Introduction}

Radiative heat waves (Marshak waves) are a basic phenomena in high energy density physics (HEDP) laboratory astrophysics~\cite{lindl2004}, and in modeling of astrophysics phenomena (e.g. supernova)~\cite{castor2004,zeldovich}. Specifically, once a drive laser or other energy source is applied to a sample, a radiative subsonic heat wave generates an ablative shock wave, propagating in the material in front of the heat wave. This is the basic physical process which occurs inside the walls of a hohlraum used to convert laser light into x-rays in the indirect drive approach of inertial confinement fusion (ICF)~\cite{lindl2004,rosen1996}.

The heat conduction mechanism in these high temperatures (100-300eV and higher) and opaque regions is radiation heat conduction, rather than the electron heat conduction. This radiation-dominated heat conduction mechanism occurs even though the radiation energy (or the radiation heat-capacity) and the radiation pressure themselves are negligible relative to the material energy and pressure. The wave propagates mainly through absorption and black-body emission processes (Thomson scattering is negligible in the range of 100-500eV, compared to opacity). Although the equation which correctly describes  the photons motion is the Boltzmann equation for radiation~\cite{pombook}, when the radiation is close to local thermodynamic equilibrium (LTE), the angular distribution of the photons is close to be isotropic and diffusion approximation yields a very good description of the exact behavior. The frequency distribution is close to a Planckian with the same temperature of the material. In this case, the governing equation is replaced by a simple single temperature conduction diffusive equation, where the diffusion equation is determined by the Rosseland mean opacity~\cite{heizler,zeldovich,hr}.

Roughly, Marshak waves can be subdivided to supersonic waves, and subsonic waves. When the wave propagates faster than the sound velocity of the material, the material hydrodynamics motion is negligible and the Marshak wave is considered to be {\em supersonic}. When the wave propagates slower than the sound velocity, hydrodynamics should be taken into account and the radiation conduction equation is solved as part of the energy conservation equation of the hydrodynamics system of equations~\cite{pombook}. This is the {\em subsonic} Marshak wave. In this case, a strong ablation occurs, causing the heated surface to rapidly expand backwards. Due to momentum conservation, a strong shock wave starts to propagate from the heat wave front (the ablation front), and it propagates faster than the heat front itself.

Marshak offered a self-similar solution to the supersonic region, in the case that the material's opacity and heat capacity can be described through simple power-laws~\cite{marshak}.
Later, a self-similar solution for the subsonic case was introduced, for the hydrodynamics equations coupled to the radiation conduction equation~\cite{ps,ps2,ger3}. Many solutions based on self-similar solutions or perturbation theories were offered, backed also by direct simulations~\cite{rosenScale2,rosenScale3,hr,garnier}. Those solutions were recently used to analyze both qualitatively and quantitatively supersonic Marshak wave experiments~\cite{avner1,avner2}.
We note that the self-similar solutions include only the heat wave region itself, but not the shock region, since the whole subsonic motion is not self-similar altogether.

Recently, Shussman et al. offered a full self-similar solution to the subsonic problem for a general power-law dependency of the temperature boundary condition (BC), based on patching two self similar solutions, each valid for a different region of the problem~\cite{ts1}. Shussman et al. used the Pakula \& Sigel self similar solution~\cite{ps,ps2,ger3} for the heat region, which determines a power-law time-dependent pressure BC for the shock region, and strong shock Hugoniot relations for the other BC. Since the full solution is composed of two regions with different physical regimes: heat region ($\approx 100$eV), and shock region ($\approx 1-10$eV), a further important step was the implementation of different equation-of-states (EOS) (a binary-EOS) for the two regions~\cite{ts2}. These works have yielded analytical expressions for the hydrodynamic parameters such as the ablation pressure, and the temperature dependent (power-law) shock velocity boundary condition.

The recent progress described above is the basis for an {\em analytical} re-visit of the experiments aimed for the evaluation of radiation temperature drive in hohlraums, by measurement of the shock velocity inside a wedged well characterized aluminium sample attached to the hohlraum wall, first presented by Kauffman et al.~\cite{kauff1,kauff2}.
The relation between the hohlraum radiation temperature and the shock velocity, is a scaling analytical fit to full radiation-hydrodynamic simulations. Based on newly performed simulations, this scaling relation was argued to be a non-universal, and specifically to depend on the laser pulse duration~\cite{sini1,sini2}. In this study we {\em analytically} examine the sensitivity of the radiation temperature to shock velocity scaling relation, to the different parameters of the temperature profile, such as the temperature profile's duration and its temporal shape, not just qualitatively but in fully quantitative manner. We use Shussman et al. analytical model~\cite{ts1} for the ablation shock region, and take advantage of the binary EOS model which is of most importance in modeling of these experiments~\cite{ts2}.

The present paper is structured in the following manner: first, in Sec.~\ref{eperiments}, we will review previous experiments and scaling-laws offered to determine the radiation temperature through the shock velocity measurement in aluminium. In Sec.~\ref{model} we derive the self-similar semi-analytic equations that determine the explicit dependency of the resulting drive temperature in the different profiles' parameters. Next, in Sec.~\ref{results} the model results are presented, showing the dependency of the results on the various parameters, and reproducing the experiments and simulations presented in the literature. The model results will explain the differences between the various experiments' results. A short discussion is presented in Sec. \ref{discussion}.

\section{The Shock-wave Measurements experiments and the different scaling laws}
\label{eperiments}

\begin{figure}
\centering{
(a)
\includegraphics*[width=8.4cm]{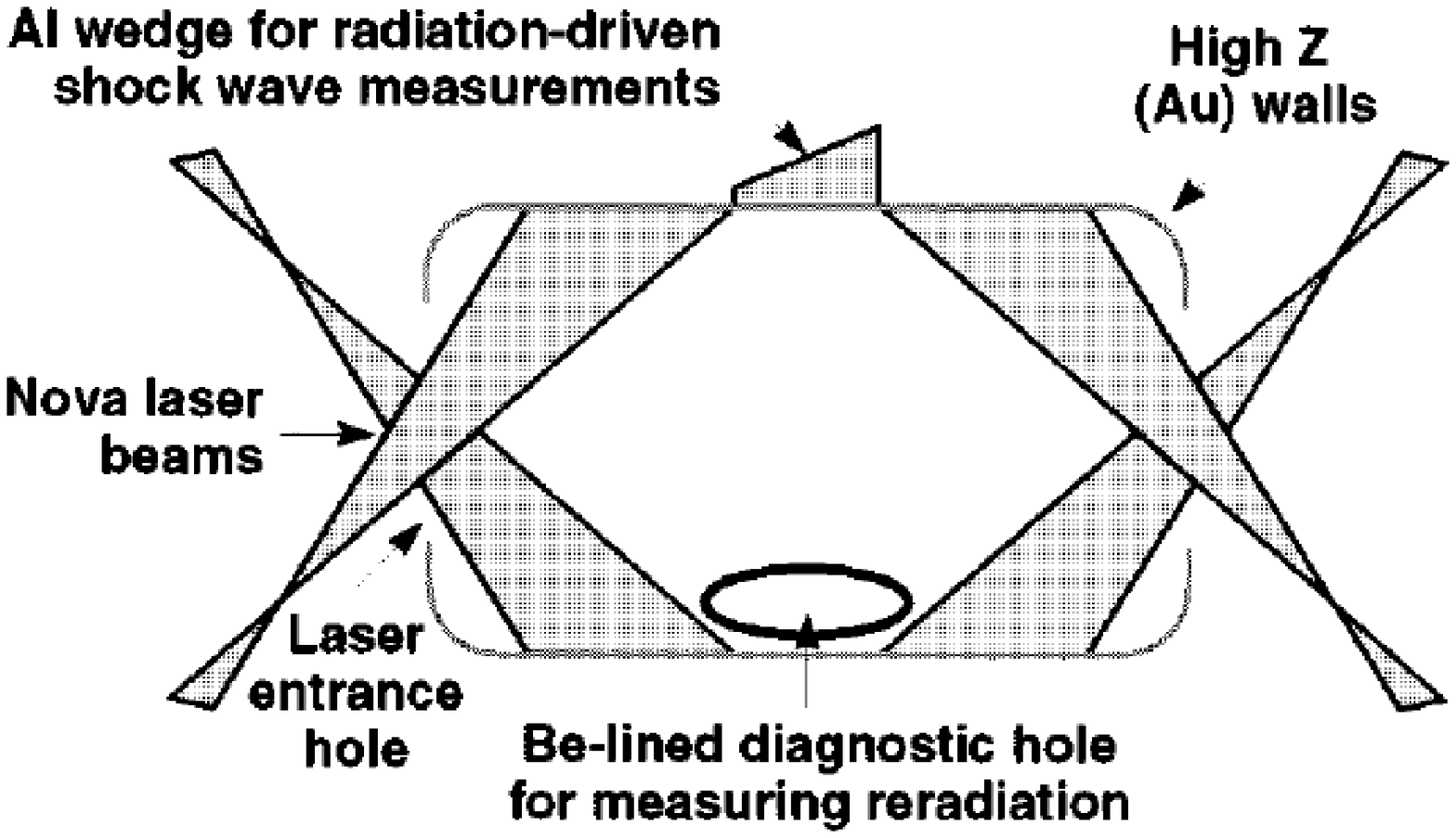}
(b)
\includegraphics*[width=6.6cm]{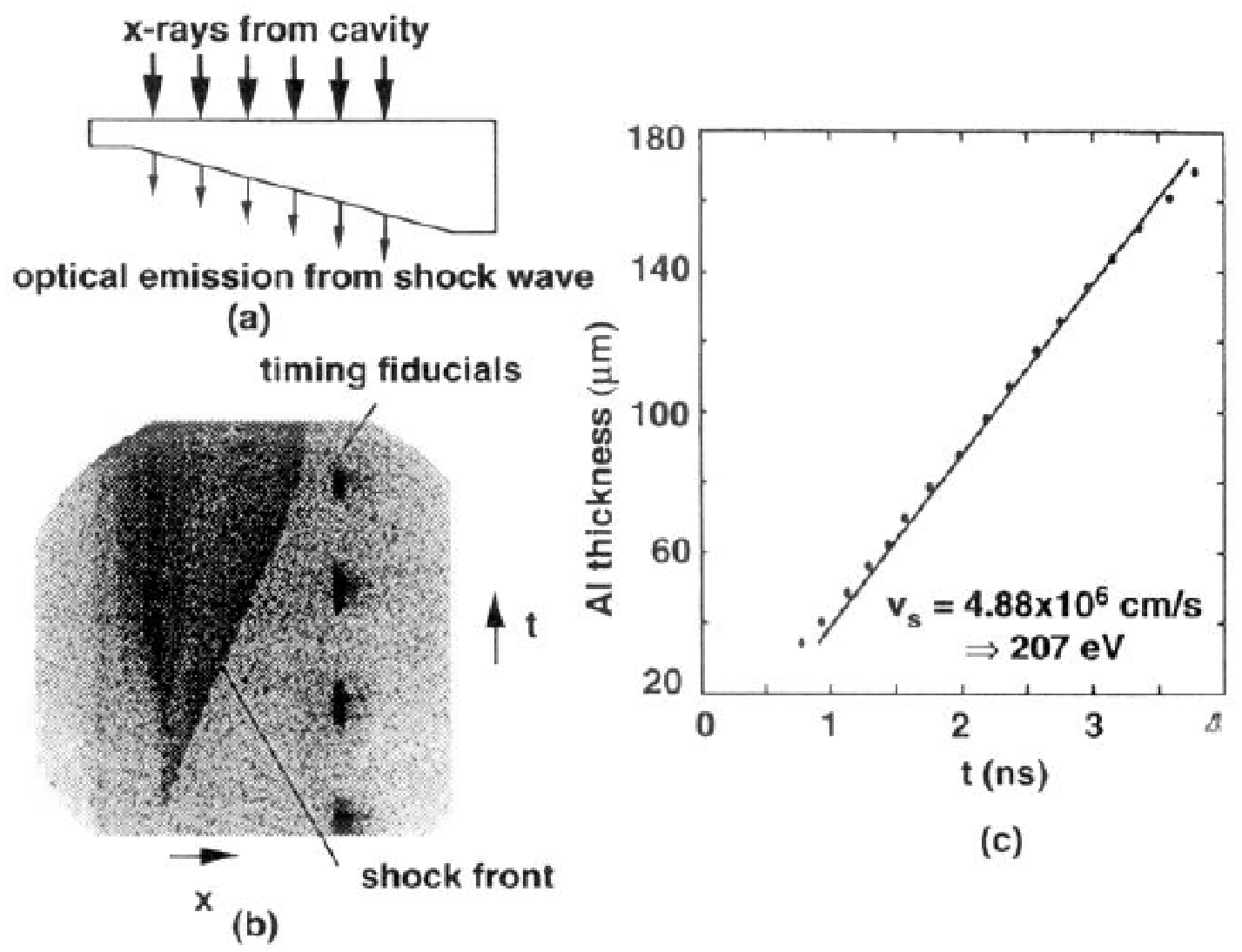}
}
\caption{(a) Typical experimental setup of evaluation hohlraum radiation temperature through the measurement of the shock wave velocity in aluminium wedge. (b) The shock velocity can be solved from the shock position, which is measured as a function of time, due to the varying sample thickness. Reproduced with permission from \prl~73, 2320  (1994). Copyright 1994 American Physical Society.}.
\label{exp_set}
\end{figure}

The experimental method for the evaluation of the radiation drive temperature by measuring the shock velocity was proposed by Hatchett et al.~\cite{hatchett,campbell,amer_prl} and performed by Livermore groups~\cite{kauff1,kauff2}, and right after that by the German group~\cite{german_prl,german}. A schematic diagram of the American experiments can be seen in Fig.~\ref{exp_set}(a), where high energy laser beams enter a hohlraum to generate a high-temperature x-ray cavity of 100-300eV.

The laser energy is absorbed in the high-$Z$ material hohlraum walls, and generates soft x-rays which undergo thermalization inside the hohlraum. The hohlraum walls are made of high-$Z$ optically thick materials (usually gold) to achieve a large laser to x-ray conversion efficiently. Since the hohlraum walls have a finite opacity, a nonlinear radiative heat wave is generated and quickly becomes subsonic. A diagnostic hole is covered by a wedged sample made of reference-material which should be well characterized (by means of opacity and EOS) so aluminium is the natural choice. Although aluminium is less opaque than gold to x-rays due to its relatively low-$Z$, it is opaque enough so that the heat wave inside the aluminium is subsonic as well. The high energy ablates the inner surface of the hohlraum walls and the aluminium wedge, yielding an ablative density profile. As a consequence (due to conservation of momentum), an ablative (radiation-driven) shock wave is propagating in front of the heat wave. The wedged shape allows to temporally resolve the position of the shock (Fig.~\ref{exp_set}(b)), and the shock velocity is determined from the slope. In the German version experiments, a series of targets with different thicknesses were used instead of the wedge, so the shock velocity can be determined with somewhat less accuracy and temporal resolution~\cite{german}.

\begin{figure}
\centering{
\includegraphics*[width=8cm]{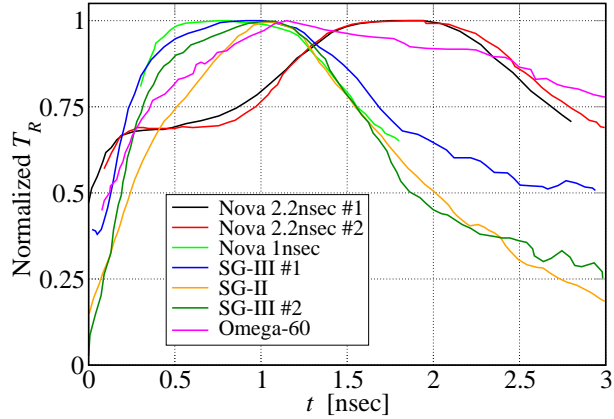}
}
\caption{Different radiation temperature profiles that were used in the various experiments. Nova 2.2nsec \#1 pulse was taken from~\cite{kauff1}, Nova 2.2nsec \#2 and Nova 1nsec pulses were taken from~\cite{kauff2}. The SG (ShenGuang) pulses were taken from~\cite{sini1}, whereas the Omega-60 pulse was taken from~\cite{back2}. The profiles are divided roughly to short pulses of $\approx1$nsec and long pulses of$\approx2$nsec. The temperatures were normalized to their maximal value.
}
\label{exp_pul}
\end{figure}

Using the well-known material properties of aluminium, a scaling fit can be determined by calibrating exact simulations, to formulate the relation between the incoming radiation temperature and the out-going shock velocity~\cite{kauff1,german}. Kauffman's scaling was fitted to the Nova experiments~\cite{kauff1,kauff2}:
\begin{equation}
T_{\mathrm{Rad}}^{\mathrm{Kauffman}}=0.178\cdot D_S^{0.63}
\label{koyfman}
\end{equation}
Where the shock velocity $D_S$ is measured in km/sec, and the drive radiation temperature $T_{\mathrm{Rad}}$ is in heV ($=100$eV). The power-law form is supported by self-similar analysis~\cite{hatchett}. This scaling relation was calibrated for relatively high-temperature, $200<T_{\mathrm{Rad}}<300$eV, and mostly for Nova facility long-pulses duration, $2-2.5$nsec (see Fig.~\ref{exp_pul}, but short pulses were examined through this scaling relation as well). We note that Remington et al. have used this method to measure the drive temperature in longer pulses (3nsec) for Rayleigh--Taylor experiments~\cite{remington}.

Eidmann et al. have set different scaling relation that covers the low-temperature range $100<T_{\mathrm{Rad}}<150$eV and short-pulse duration, 0.8nsec in the Gekko-XII facility~\cite{german}:
\begin{equation}
T_{\mathrm{Rad}}^{\mathrm{Eidmann}}=0.184\cdot D_S^{0.6}
\label{eidmann}
\end{equation}

A decade ago, there was a renewed interest in these experiments and Kauffman's scaling relation, due to the works by Li et al.~\cite{sini1,sini2}. Following direct full simulations, they claim that Kauffman's scaling relation is not universal, but rather is only correct for Nova long laser pulses ($\approx$2-2.5nsec), in which case their simulations reproduce quantitatively the Nova experiment (also in the work of~\cite{india_old}). In shorter pulses of $\approx$1nsec, like in the SG-II and SG-III (ShenGuang) facilities, a new scaling relation is introduced~\cite{sini1}:
\begin{equation}
T_{\mathrm{Rad}}^{\mathrm{Li}}=0.1579\cdot D_S^{0.647}
\label{li}
\end{equation}
Li et al. claim that the difference between the two scaling relations is due to the different temperature profile's duration, while the dependency on the temporal shape is negligible. {\bf We will examine these two claims carefully in this study}.

Quite recently, a new scaling-relation was offed by Mishra et al., based on new simulations~\cite{india} using a modified version of the widely-used MULTI code~\cite{ramis}. The simulations were carried out using a {\em constant radiation temperature} boundary condition with a {\em long pulse} of 3nsec, and temperature range of $100<T_{\mathrm{Rad}}<500$eV. The paper shows a transition of the slope in the $T_{\mathrm{Rad}}(D_S)$ curve in $T_{\mathrm{Rad}}\approx275$eV, however the high-range is beyond the experiments regime. Below $T_{\mathrm{Rad}}\approx275$eV, the scaling relation is fitted to~\cite{india}\footnote[2]{The exact power in the scaling law which appears in~\cite{india} is 0.65. However, it does not fit Ref.~\cite{india} own simulations data, so we assume they have used two-digits round. The exact value that fits their simulations data (keeping the pre-factor unchanged) is 0.653, as in Eq.~\ref{mishra}.}:
\begin{equation}
T_{\mathrm{Rad}}^{\mathrm{Mishra}}=0.1565\cdot D_S^{0.653}
\label{mishra}
\end{equation}
which surprisingly is closer to Li short-pulse scaling relation Eq.~\ref{li} rather than to Kauffman's long-pulse scaling relation Eq.~\ref{koyfman}. This fact will also be explained through our analytic model. Mishra et al. offer scaling laws fits for high-$Z$ material as well, however, the fit is affected again, mostly by the high-temperature range $250<T_{\mathrm{Rad}}<500$eV. Das et al.~\cite{Das} have set a new set of simulations for testing a new opacity code. The simulations' results lie closer to Li~\cite{sini1} results than to Kauffman's~\cite{kauff1} in most of the examined regime, however, it is unclear which boundary condition/temperature profile's shape and duration were used to perform these simulations.

We note that in some of the experiments, the radiation temperature results were compared to direct measurement of the x-rays emitted from the hohlraum walls, using x-ray diodes (XRD) array~\cite{kauff1,kauff2} or transmission grating spectroscopy (TGS)~\cite{german}. This measurement method allows to follow the temporal shape of the radiation temperature (see. Fig.~\ref{exp_pul}). The agreement between the two methods was quite good, though as we will see later (Sec.~\ref{Td_sec}) there is a noticeable difference between the different radiation temperatures; the temperature of the incoming x-rays (drive) is higher than the observed wall temperature~\cite{MordiLec,MordiPoster} (see also~\cite{avner1,avner2}, for the importance of this effect in implementation of modeling supersonic Marshak wave).

In Fig.~\ref{exp_pul} we present several typical radiation temperature profiles as measured using the XRD/TGS techniques, that were investigated in the various studies. As will be shown later, from these measured profiles we can limit the regime of the temperature profile's temporal-behavior, such as temperature profile's duration and temporal shape, for the use of our self-similar solution. Nova's short pulse (1nsec) is quite flat ($T_R\sim t^{0}$) after its rise-time, whereas the longer pulses (2-2.5nsec) have two (flat) steps structure~\cite{kauff1,kauff2}. The SG-III pulses rises as $T_R\sim t^{0.1-0.15}$ after short rise-time, while The SG-II typical pulse rises slower, $T_R\sim t^{0.3-0.35}$~\cite{sini1}. Some pulses, like the one that was used in the Back's et al. supersonic Marshak waves experiments~\cite{back2} decrease with time, $T_R\sim t^{-0.1}$ (1-2.5nsec) after its first 1nsec rise-time. {\bf In this study we exploit the general power-law solution of~\cite{ts1,ts2} to study the sensitivity of the $T_R(D_S)$ curves to the properties of the temperature profiles' parameters.} 

It should be noted that although most of the works investigating the radiation drive temperature used aluminium, another group measured the shock-velocity in quartz~\cite{quartz}. The empirical scaling law ($T_{\mathrm{Rad}}^{\mathrm{quartz}}=0.214D_S^{0.57}$) is of course different than the aluminium scaling-laws. We shall not discuss this work here, since the focus of the present work is to explain the differences between the different scaling-laws using aluminium. 

\section{The (semi-) analytic Model}
\label{model}
In this section we present the semi-analytic model for estimating the $T_R(D_S)$ curves. The derivation is presented for aluminium, with a couple of delicate issues that have to be done carefully for yielding accurate quantitative results: The detailed EOS, and the calibration of opacity factors in order to include transport effects. The procedure is as follows:
\begin{itemize}
       \item Solving semi-analytically the ablative heat region as a function of the surface temperature BC ($T_W$) which produces an analytic expression for the ablation pressure (see Sec.~\ref{heat_sec}). This procedure involves the use of an opacity factor which is calibrated from an exact Monte-Carlo simulations (see Sec.~\ref{opac_sec}).
       \item Solving semi-analytically the shock region as a function of the ablation pressure which produces an analytic expression for the resulting shock velocity $D_S$ (see Sec.~\ref{shock_sec}). This procedure involves the use of the detailed EOS of aluminium (see Sec.~\ref{eos_sec}).
       \item Determining the drive temperature $T_R$ for the given $D_S$ from the surface temperature $T_W$ using the self-similar solution of the flux from the heat region solution (see Sec.~\ref{Td_sec}).
     \end{itemize}
The final analytic expressions of the derivation are summarized in Sec.~\ref{fin_sec}.

\subsection{The ablative heat region}
\label{heat_sec}
To establish a self-similar solution of the ablative heat region, one must assume a power-law relation of the Rosseland mean opacity $\kappa_R$ and the internal energy $e$, as a function of the temperature $T$ and the density $\rho$~\cite{ps,ts1}. We use Hammer \& Rosen notations, where the temperature has units of $\mathrm{heV}\equiv 100\mathrm{eV}$ and the density has $\mathrm{gr/cm^3}$ units~\cite{hr}:
\begin{subequations} \label{pwrlaws}
\begin{equation} \label{ross}
\frac{1}{\kappa_R}=\frac{g}{\kappa_0(t_{\mathrm{Pulse}})}T^\alpha\rho^{-\lambda}
\end{equation}
\begin{equation} \label{pwrlaw_energy}
e=fT^\beta\rho^{-\mu}
\end{equation}
\end{subequations}
$\kappa_0(t_{\mathrm{Pulse}})$ is a unitless factor which multiplies the nominal opacity. We use it here to calibrate the diffusion approximation solution to the exact transport (Boltzmann) solution using IMC simulations (the calibration is found to be a temperature profile's duration $t_{\mathrm{Pulse}}$ dependent, see Sec.~\ref{opac_sec}). This is due to the fact that aluminium is not an extremely-opaque material, so the diffusion solution yields a too fast heat-wave (unlike gold, in which case the diffusion approximation yields an excellent transport solution).

In addition, we assume an ideal gas-like EOS using an adiabatic factor $\gamma_1$, again using Hammer \& Rosen notations (the index 1 denotes the heat-region):
\begin{equation}
P(\rho,T)=r_1\rho e(\rho,T)\equiv(\gamma_1-1)\rho e(\rho,T)
\label{def_r1}
\end{equation}
where $\gamma_1\equiv(r_1+1)$ is the ideal gas parameter in the ablation region. The different parameters for aluminium, which is the material of the shock waves experiments, are given in Table~\ref{alum_params}. The EOS analytical parameters $\beta$, $\mu$, $f$ and $r_1$ {\bf for the heat region} are taken from~\cite{heat_eos}, while the opacity parameters $\alpha$, $\lambda$ and $g$ are fitted to the up-to-date opacity code CRSTA tables for the range of $1-3\mathrm{heV}$~\cite{Kurz2012,Kurz2013}.
\begin{table}[!htb]
 \centering
 \caption{\bf Power law fits for the opacity and EOS of aluminium in the temperature range of $1-3\mathrm{heV}$.}
 \label{table:pwr_law_opac_eos} 
 \begin{tabular}{|c|c|} \hline 
 \multicolumn{1}{|c|}{{\bf Physical Quantity}} &
 \multicolumn{1}{c|}{{\bf Numerical Value}} \\ \hline
 \ $f$ & $9.04$ [MJ/g] \\ \hline
 \ $\beta$ & $1.2$ \\ \hline
 \ $\mu$ & $0$ \\ \hline
 \ $g$ & $1/1487$ $[\mathrm{g/cm^2}]$ \\ \hline
 \ $\alpha$ & $3.1$ \\ \hline
 \ $\lambda$ & $0.3685$ \\ \hline
 \ $r_1\equiv(\gamma_1-1)$ & $0.3$ \\ \hline
 \end{tabular}
\label{alum_params}
\end{table}

At~\cite{ts1}, the solution of the heat region is given for a general power-law boundary condition: 
\begin{equation} \label{temperature_pwrlaw}
T_W(t)=T_0t^{\uptau}
\end{equation}
where in general $T_W(t)$ is the inner {\em surface temperature} of the sample (in heV) and $t$ is the time (in nsec). In this study $T_W$ is the inner surface of the aluminium wedge which is attached to the hohlraum hole (see Fig.~\ref{exp_set}(a), the surface at which the x-rays from the cavity hit the wedge in Fig.~\ref{exp_set}(b)). The hohlraum temperature temporal profiles, measured by the XRD/TGS diagnostics in the various experiments, set the limits of validity of $\uptau$ for our investigations to be $-0.05\leqslant\uptau\leqslant 0.3$.

Shussman \& Heizler~\cite{ts1} present a self-similar solution for the ablative pressure, located at the heat-front position $x_F$, as well as the total energy stored in the heat region (which is almost equal to the total energy, since the energy in the shock region is negligible):
\begin{subequations}
\label{heat_eqs}
\begin{equation}
P_F(t)=p_0(\uptau)\kappa_0^{P_{\omega_1}}(t_{\mathrm{Pulse}})T_0^{P_{\omega_2}}t^{\uptau_S(\uptau)}\equiv P_0(\uptau)t^{\uptau_S(\uptau)}
\kappa_0^{P_{\omega_1}}(t_{\mathrm{Pulse}}) \left[\mathrm{Mbar}\right]
\label{pressure}
\end{equation}
\begin{equation}
E_W(t)=e_0(\uptau)\kappa_0^{E_{\omega_1}}(t_{\mathrm{Pulse}})T_0^{E_{\omega_2}}t^{E_{\omega_3}(\uptau)} \left[\mathrm{\frac{hJ}{mm^2}}\right]
\label{energy}
\end{equation}
\end{subequations}
The different powers are determined from dimensional analysis while the pre-factors are determined by solving the dimensionless ODE, as derived in details in~\cite{ts1}. In~\cite{ts1} the procedure is derived for gold parameters, whereas here we present the equivalent results for aluminium, using Table,~\ref{alum_params}. The powers for the ablative pressure and energy are:
\begin{subequations}
\begin{equation}
\label{tau_shock}
\uptau_S(\uptau)=\frac{-1+\mu+(4+\alpha+\beta\lambda)\uptau-(4+\alpha)\mu\uptau}{2+\lambda-2\mu}
\end{equation}
\begin{equation}
P_{\omega_1}=-\frac{1-\mu}{2+\lambda-2\mu}\approx-0.422
\end{equation}
\begin{equation}
P_{\omega_2}=\frac{4+\alpha+\beta\lambda-(4+\alpha)\mu}{2+\lambda-2\mu}\approx3.184
\end{equation}
\begin{equation}
E_{\omega_1}=\frac{2-3\mu}{4+2\lambda-4\mu}\approx-0.422
\end{equation}
\begin{equation}
E_{\omega_2}=\frac{8+2\alpha+2\beta+3\lambda\beta-3(4+\alpha)\mu}{4+2\lambda-4\mu}\approx3.784
\end{equation}
\begin{equation}
E_{\omega_3}=\frac{2+2\lambda-\mu+(2(4+\alpha+\beta)+3\beta\lambda)\uptau-3(4+\alpha)\mu\uptau}{4+2\lambda-4\mu}
\end{equation}
\end{subequations}

\begin{figure}
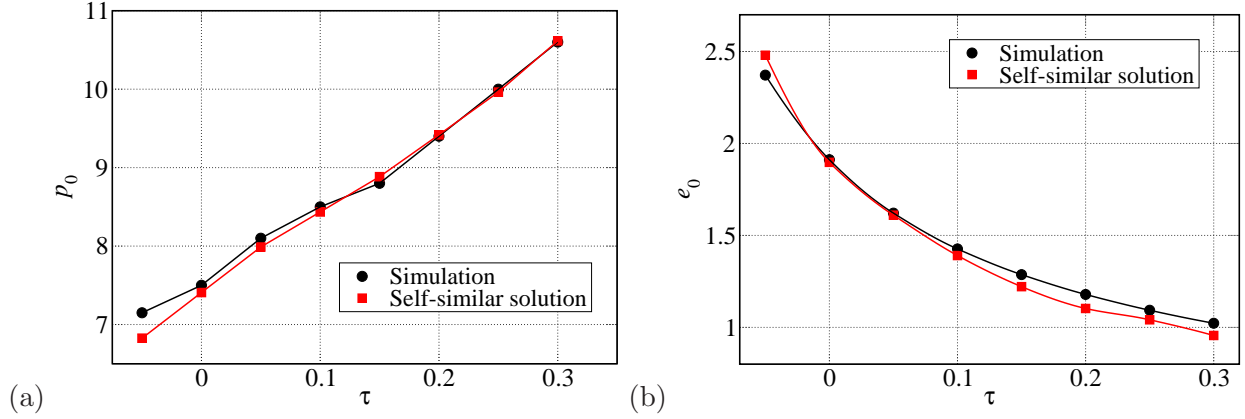

\centering{
(a)
\includegraphics*[width=7.5cm]{p0.eps}
(b)
\includegraphics*[width=7.5cm]{e0.eps}
}
\caption{(Color online) (a) The pressure parameter as a function of the temperature power dependence $\uptau$, to be used in Eq.~\ref{pressure}.
(b) The energy parameter as a function of the temperature power dependence $\uptau$, to be used in Eq.~\ref{energy}.}
\label{ss_coef}
\end{figure}

The constants pre-factors $p_0(\uptau)$ and $e_0(\uptau)$ are determined from the solution of the dimensionless ODE, and are presented in the red curves in Fig.~\ref{ss_coef}, as a function of the temperature BC $\uptau$. In addition, we have performed direct simulations for validating the numerical constant (in black curves). The simulations were performed using a one-dimensional radiative-hydrodynamics code, which couples Lagrangian hydrodynamics with implicit LTE diffusion radiative conduction scheme, in an operator-split method. The hydrodynamics code uses explicit hydrodynamics using Richtmyer's artificial viscosity and Courant's criterion for a time-step. In the diffusion conduction scheme, the time-step is defined dynamically such that the temperature will not change in each cell by more than 5\% between time steps (for more details regarding the radiative-hydrodynamics code, see~\cite{ts1,ts2,avner1,avner2}). In both schemes, we have used a converged constant space intervals. The matching between the self-similar solution and the simulations is very good.

\subsection{The shock region}
\label{shock_sec}

The database for the sock region is simply the EOS (heat conduction is negligible in this region). Again, we assume an ideal-gas EOS (the index 2 denotes the shock-region):
\begin{equation}
P(\rho,T)=r_2\rho e(\rho,T)\equiv(\gamma_2-1)\rho e(\rho,T)
\label{def_r2}
\end{equation}
where $\gamma_2\equiv(r_2+1)$ is the ideal gas parameter in the shock region. Notice that we use a binary EOS following~\cite{ts2} using $r_1\ne r_2$. As opposed to $r_1$, the determination of $r_2$ is more complex, as it is a function of the shock velocity -- $r_2(D_S)$, following the detailed Hugoniot EOS data of aluminium~\cite{hugo1,hugo2}. The values of $r_2(D_S)$ are discussed in Sec.~\ref{eos_sec}.

\begin{figure}
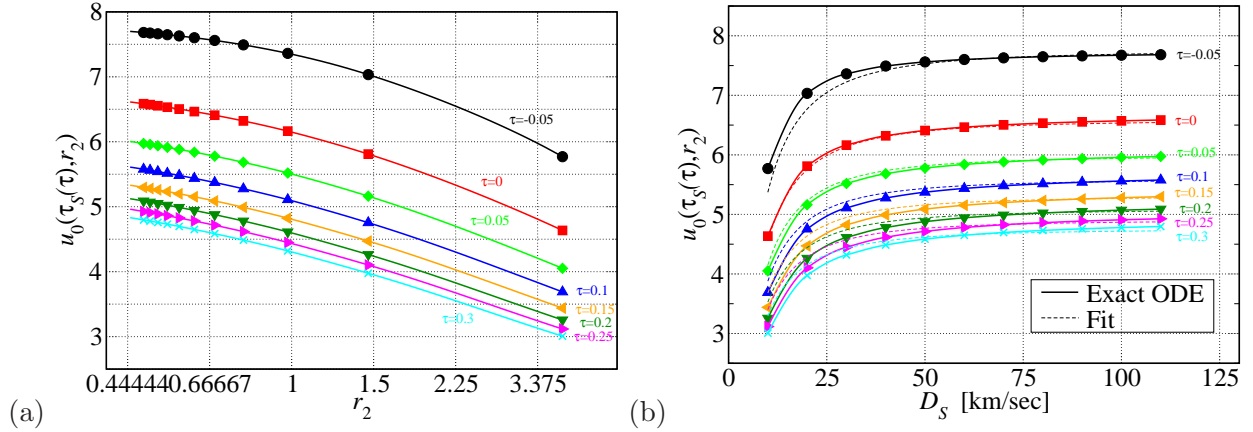

\centering{
(a)
\includegraphics*[width=7.5cm]{u0_2.eps}
(b)
\includegraphics*[width=7.5cm]{u0_1.eps}
}
\caption{(Color online) (a) The velocity parameter as a function of the EOS parameter $r_2(D_S)$ for different $\uptau$ (via $\uptau_S$), to be used in Eq.~\ref{shock_T0_quant}. (b) The velocity parameter as a function of the shock velocity $D_S$ for different $\uptau$ (solid curves). A simple fit for these curves is sown in the dotted curves through Eq.~\ref{fit} with a maximal error of 10\% (above $D_S=30$km/sec its maximal error is 2.5\%).}
\label{ss_coef2}
\end{figure}

Following~\cite{ts1,ts2}, we take the ablation pressure achieved from the heat region, Eq.~\ref{pressure} as a BC for the shock region:
\begin{equation} \label{power_pwrlaw}
P(t)=P_0t^{\uptau_S}
\end{equation}
where $P_0=p_0(\uptau)\kappa_0^{P_{\omega_1}}(t_{\mathrm{Pulse}})T_0^{P_{\omega_2}}$ and $\uptau_S$ is defined by Eq.~\ref{tau_shock}. Both $P_0$ and $\uptau_S$, are known functions of $\uptau$, which determines the shape of the temperature profile (Eq.~\ref{temperature_pwrlaw}). The second BCs are taken to be the strong shock limit of the Hugoniot relations. Shussman et al. present self-similar solution for the particle and shock velocities, located in the shock-front position $x_S$ of this form:
\begin{subequations}
\label{shock_T0_quant}
\begin{equation}
u_S(t)=u_0(\uptau_S(\uptau),r_2(D_S))P_0^{\frac{1}{2}}(\uptau)t^{\frac{\uptau_S(\uptau)}{2}} \left[\mathrm{km/sec}\right]
\label{sub_T0_quant2}
\end{equation}
\begin{equation}
D_S(t)=\frac{r_2(D_S(t))+2}{2}u_S(t)
\end{equation}
\end{subequations}
The powers in Eqs.~\ref{shock_T0_quant} are determined by a dimensional analysis, and the pre-factor $u_0(\uptau_S(\uptau),r_2(D_S))$ by the solution of the dimensionless ODE and is given in Fig.~\ref{ss_coef2}.
We can see in Fig.~\ref{ss_coef2}(a) the dependency of $u_0$ on both $r_2$ and $\uptau$ (via $\uptau_S$). The dependency of $u_0$ on $\uptau$ decreases for $\uptau>0.2$. Plotting $u_0(D_S)$ explicitly in Fig.~\ref{ss_coef2}(b) (through the $r_2(D_S)$ functional form) discovers
that except for low shock velocities of $D_S<30$km/sec, the value of $u_0(D_S)$ increases slowly. We have also performed a simple fit to the exact curves of $u_0(D_S)$ in Fig.~\ref{ss_coef2}(b), which assumes a separation of variables, $f(D_S)$ and $g(\uptau)$. Such a ``universal solution" would make an easy-to-use resource to researchers for future work. The fit has the form:
\begin{equation}
u_0(D_S)\approx f(D_S)\cdot g(\uptau)=\left(6.61-\frac{1035}{(D_S+9.2)^{2.11}}\right)\cdot \frac{0.58}{(\uptau+0.1)^{0.235}}
\label{fit}
\end{equation}
The fit of Eq.~\ref{fit} is shown in the dotted curves in Fig.~\ref{ss_coef2}(b), and has an accuracy of 10\% (above $D_S=30$km/sec its maximal error is 2.5\%). Such an error represents a maximal error of 2eV in the $T_{\mathrm{Rad}}(D_S)$ curves (which is negligible for any practical use), that are shown later in Sec.~\ref{results} (Fig.~\ref{results2}). A fit for the $u_0(r_2)$ curves that are shown in Fig.~\ref{ss_coef2}(a) can also be derived by just substituting the simple analytic relation of $r_2(D_S)$ (see later Eqs.~\ref{r2_final} and~\ref{r_bin1}) in the fit of Eq.~\ref{fit}. We note that in this work we've used the exact values of $u_0(D_S)$ self-similar solutions and not the approximated fit.

Finally, substituting Eq.~\ref{pressure} in Eqs.~\ref{shock_T0_quant} yields the dependency of the out-going shock velocity in the shock region, as a function of the {\em surface temperature} of the heat region (the final analytical relation between $D_S$ and the drive temperature $T_{\mathrm{Rad}}$ will be discussed in Sec.~\ref{Td_sec}.):
\begin{equation} \label{final_shock}
D_S(t_{\mathrm{Pulse}})=\frac{r_2(D_S)+2}{2}u_0(\uptau_S(\uptau),r_2(D_S))\sqrt{p_0(\uptau)\kappa_0^{P_{\omega_1}}(t_{\mathrm{Pulse}})}T_0^{\frac{P_{\omega_2}}{2}}t^{\frac{\uptau_S(\uptau)}{2}}_{\mathrm{Pulse}}
\end{equation}
Note that Eqs.~\ref{final_shock} and~\ref{shock_T0_quant} are nonlinear in $D_S$ due to the dependency of $r_2(D_S)$. This relation is calculated from the up-to-date detailed data of Hugoniot relation for Al~\cite{hugo1,hugo2}, and will be discussed in Sec.~\ref{eos_sec}.

\subsection{Calibration of the opacity factor $\kappa_0(t_{\mathrm{Pulse}})$}
\label{opac_sec}

There is still one delicate issue, concerning the self-similar solution of the heat region, for determining the ablative pressure and the stored energy. In Eq.~\ref{heat_eqs} we specify the dependency of the physical parameters in $\kappa_0$, an arbitrary opacity multiplier of the nominal opacity in Eq.~\ref{ross}.

We use this parameter as a tool to set radiative transport effects corrections, due to the relatively low-opacity of aluminium ($Z=13$).
Although the heat wave in aluminium in these experiments in fully subsonic, and it generates the strong-shock limit in the shock region, it is still more optically-thin than the heat wave in high-$Z$ materials, such as gold. In high-$Z$ materials, the heat wave is well modeled using the LTE diffusion approximation. In the (nominal-opacity) aluminium case, diffusion yields a heat front which is too fast comparing to the exact transport solution. We test this difference by a series of gray implicit-Monte-Carlo (IMC)~\cite{IMC} simulations that we set in our transport code with nominal opacities and EOS, as in Eq.~\ref{pwrlaws} and~\ref{def_r1}. 
The IMC simulations have used one-dimensional radiative-hydrodynamics code, which couples Lagrangian hydrodynamics (same code that was used in Sec.~\ref{heat_sec}) with 1D IMC (Fleck \& Cummings) scheme~\cite{IMC} in operator-split method, while the diffusion calculations couple the hydrodynamics to implicit LTE diffusion radiative conduction instead (again, for more details regarding the radiative-hydrodynamics code, see~\cite{ts1,ts2,avner1,avner2}).

\begin{figure}
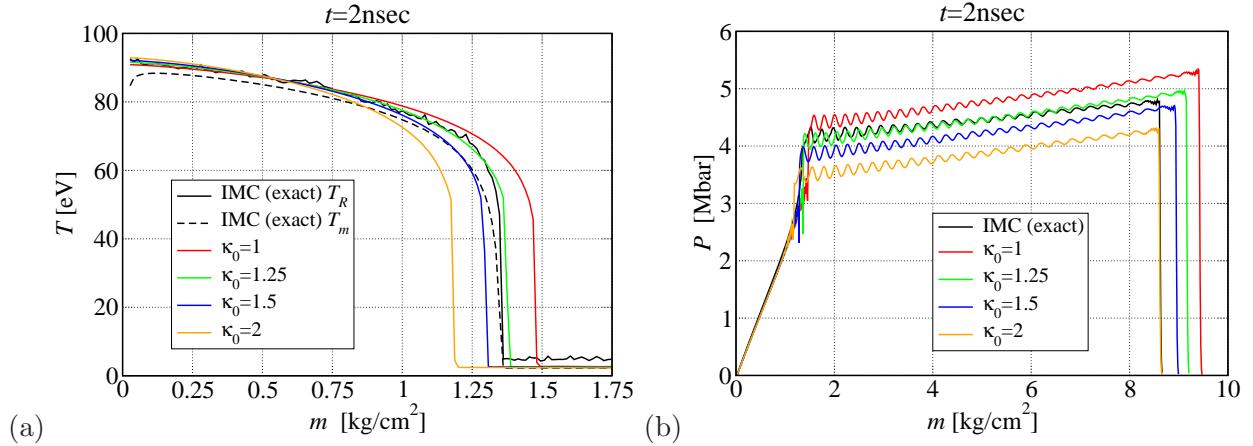

\centering{
(a)
\includegraphics*[width=7.7cm]{Tkappa.eps}
(b)
\includegraphics*[width=7.3cm]{Pkappa.eps}
}
\caption{(a) The temperature profiles (both material and radiation) using exact IMC simulation of heat wave in aluminium using $T_{\mathrm{Rad}}=100$eV in $t=2$nsec, and with LTE diffusion approximation using different opacity multipliers $\kappa_0$. (b) Same with the pressure profiles.}
\label{calib1}
\end{figure}

\begin{figure}
\centering{
\includegraphics*[width=7.5cm]{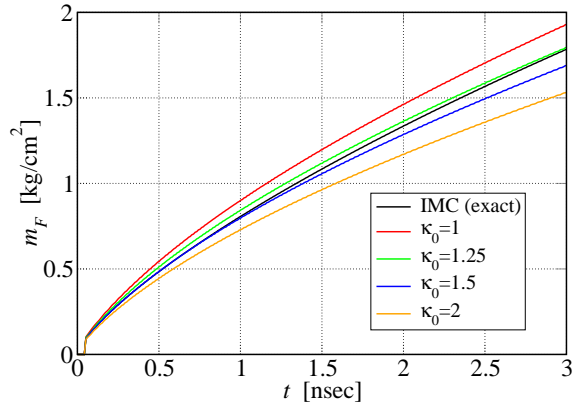}
}
\caption{The heat front position $x_F$ as a function of the time using exact IMC simulation of heat wave in aluminium using $T_{\mathrm{Rad}}=100$eV, and with LTE diffusion approximation using different opacity multipliers $\kappa_0$.}
\label{calib2}
\end{figure}

In Fig.~\ref{calib1}(a) the black curves are the temperature profiles (both material and radiation) of the exact transport solution using gray IMC code with BC of $T_{\mathrm{Rad}}=100$eV in $t=2$nsec, along with LTE diffusion approximation solution (red curve), whereas in Fig.~\ref{calib1}(b) we present the pressure curves. In Fig.~\ref{calib2} we present the heat front position $x_F$ as a function of time using BC of $T_{\mathrm{Rad}}=100$eV. We can see that in both Figures, the diffusion approximation yields a too fast heat wave, as expected, and a too high ablation pressure. A possible solution is to use Flux-Limited diffusion~\cite{avner1}, however, the nonlinear diffusion coefficient prevents a self-similar solution, which is detrimental for this study.

Thus, we take advantage of the possibility to include an opacity factor multiplier in the frame of a self similar solution~\cite{ts1}, to calibrate the LTE diffusion solution to the exact transport behavior. Again, for high-$Z$ materials, where LTE yields excellent transport solution, we set $\kappa_0=1$. For Al ($Z=13$) we have performed a set of LTE diffusion simulation using different opacity multipliers in the range $1\leqslant\kappa_0\leqslant2$. We can see, for example in Fig.~\ref{calib1} that in $t=2nsec$, the LTE diffusion approximation using $\kappa_0=1.25$, matches both the correct heat front and ablation pressure (also in Fig.~\ref{calib2}, where the green curve is closer to the black IMC curve). For $t=1nsec$, the most appropriate value for $\kappa_0$ slightly varies to $\kappa_0=1.5$ (see Fig.~\ref{calib2}, whereas the blue curve is closer to the black IMC curve). In general, we find the following simple calibration relation between $\kappa_0$ and $t_{\mathrm{Pulse}}$:
\begin{equation}
\label{calib}
\kappa_0(t_{\mathrm{Pulse}})\approx 1.75-0.25t_{\mathrm{Pulse}}.
\end{equation}
It is important to note that Eq.~\ref{calib} was found by simulations to be almost universal for the drive temperature range of $100\leqslant T_{\mathrm{Rad}}\leqslant 300$eV. This calibration procedure had to be done once, using the relation of Eq.~\ref{calib} for future analysis.

\subsection{The EOS parameter $r_2(D_S)$}
\label{eos_sec}

\begin{figure}
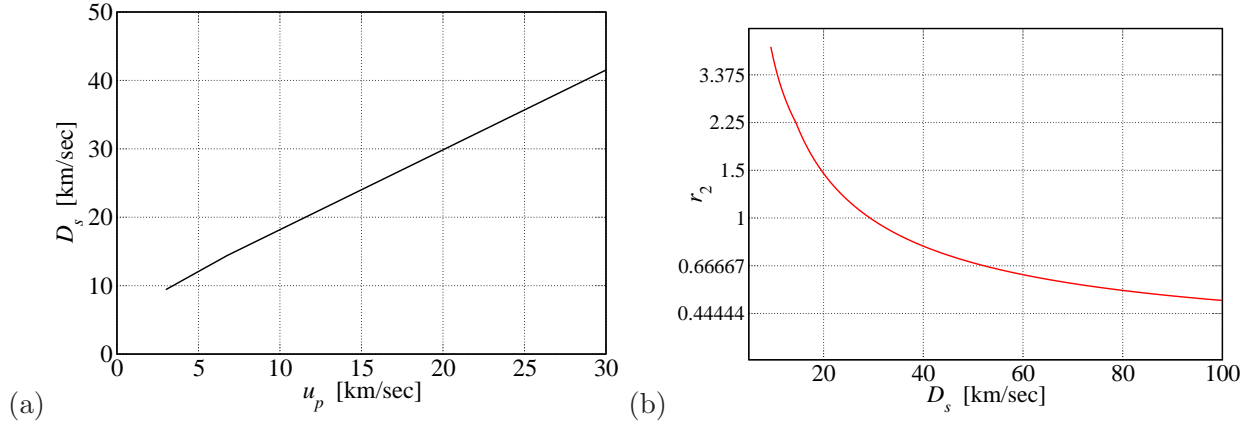

\centering{
(a)
\includegraphics*[width=7.5cm]{eos1.eps}
(b)
\includegraphics*[width=7.5cm]{eos2.eps}
}
\caption{(Color online) (a) The detailed Hugoniot EOS relation of aluminium, taken from~\cite{hugo1}. (b) The effective EOS parameter that is taken in Eq.~\ref{final_shock}, assuming strong shock relation (Eq.~\ref{strong}) using Eq.~\ref{r2_final}.}
\label{eos_plot}
\end{figure}

The detailed EOS for aluminium, and the detailed Hugoniot relations $D_S(u_p)$ (often called also $U_s-U_p$) are not the main interest of this study. However, these relations are extremely important for yielding good quantitative results using the self similar solution in the shock region (see Sec.~\ref{shock_sec}). Specifically, we are interested in the relatively high shock velocity regimes - $10\leqslant D_S \leqslant 100$km/sec. The $D_S(u_p)$ curve is often approximated by the linear relation:
\begin{equation}
\label{r_bin0}
D_S= c_0+Su_p  .
\end{equation}
We have used the values for Al from~\cite{hugo1} with an accuracy of $\approx 2-3\%$ (see also~\cite{hugo2}):
\begin{equation}
\label{r_bin1}
D_S[\mathrm{km/sec}]=\begin{cases} 5.448+1.324u_p & u_p\leqslant 6.763\mathrm{km/sec} \\
6.511+1.167u_p & u_p> 6.763\mathrm{km/sec}
\end{cases}
\end{equation}
The relation $D_S(u_p)$ for aluminium, Eq.~\ref{r_bin1}, is plotted in Fig.~\ref{eos_plot}(a). We can notice the change of the slope at $6.763\mathrm{km/sec}$.

However, The strong shock limit contradicts Eq.~\ref{r_bin0} (or Eq.~\ref{r_bin1}):
\begin{subequations}
\begin{equation}
\frac{V_S}{V_0}=\frac{r_2}{r_2+2}=\frac{\gamma_2-1}{\gamma_2+1}
\label{strong}
\end{equation}
\begin{equation}
D_S=\frac{r_2+2}{2}u_p=\frac{\gamma_2+1}{2}u_p
\end{equation}
\label{strong2}
\end{subequations}
i.e., the strong-shock relation yields $S=(\gamma_2+1)/2$ and $c_0=0$. Note that when $u_p\to\infty$, $c_0$ is negligible, and Eq.~\ref{r_bin0} tends to Eqs.~\ref{strong2}. In our case, the shock {\em is not} strong enough for the constant $c_0$ to be negligible. Thus, we define a {\em functional form} of $S'(D_S)$ in the following way:
\begin{equation}
S'(D_S)\equiv\frac{D_S}{u_p}=\frac{c_0+Su_p}{u_p},
\end{equation}
which sets an EOS parameter that is a function of $D_S$:
\begin{equation}
r'_2(D_S)=\gamma'_2-1=2\left(S'(D_S)-1\right).
\label{r2_final}
\end{equation}
Eq.~\ref{r2_final} with aluminium parameters (Eq.~\ref{r_bin1}) is presented in Fig.~\ref{eos_plot}(b) in logarithmic scale. We can see the decrease of $r_2$ with $D_S$ which softens at $D_S>20$km/sec. The resulting range of $0.45\leqslant r_2\leqslant 4$ due to the $10\leqslant D_S \leqslant 100$km/sec regime determines the limits of the tested shock region in Sec.~\ref{shock_sec} (see Fig.~\ref{ss_coef2}).

\subsection{The hohlraum temperature $T_{\mathrm{Rad}}$}
\label{Td_sec}

Eq.~\ref{final_shock} presents the nonlinear relation between the inner surface temperature (via $T_W=T_0t^{\uptau}$) and the out-going shock velocity. However, we need to associate it to the hohlraum drive temperature, which is higher. We can recognize three different radiation temperatures~\cite{MordiLec,MordiPoster}: The drive temperature $T_{\mathrm{Rad}}$, which is the temperature that characterized the incident flux toward the hohlraum's wall, the wall surface temperature $T_W$ as mentioned before, and the temperature of the emitted flux $T_{\mathrm{obs}}$, that an x-ray detector would measure, which is approximately the temperature 1mfp inside the sample (see also in~\cite{avner1,avner2}). We note also that studies which have simultaneously measured the hohlraum temperature by both the shock velocity (which should represent $T_{\mathrm{Rad}}$) and  the radiated flux (which should represents $T_{\mathrm{obs}}$) methods, have yielded different temperatures (see Table I in \cite{german}), with $T_{\mathrm{Rad}}>T_{\mathrm{obs}}$ in most measurements.

We are interested in $T_{\mathrm{Rad}}$ which characterizes the incident time-dependent flux $F_{\mathrm{inc}}(0,t)$ on the inner surface of the sample. Since this study is restricted to LTE diffusion approximation, we follow the Marshak boundary condition (an angular-integrated approximated version of the Milne boundary condition), which is defined by an integral over the incident flux~\cite{pombook,MordiLec,MordiPoster}:
\begin{equation}
F_{\mathrm{inc}}(0,t)=\frac{1}{2}\int_0^1I(\mu,0,t)\mu d\mu\equiv \sigma_{\mathrm{sb}} T_{\mathrm{Rad}}^4(t)
\label{inc_F1}
\end{equation}
were $I(\mu,0,t)$ is the specific intensity on the inner surface of the sample, $\mu$ is the cosine of the photons direction with respect to the sample's axis and $\sigma_{\mathrm{sb}}$ is the Stefan-Boltzmann constant. In the diffusion approximation, the specific intensity is a sum of its first two moments:
\begin{equation}
I(\mu,0,t)\approx cE(0,t)+3\mu F(0,t)
\label{intensity}
\end{equation}
where $E(0,t)$ is the energy density, and $F(0,t)$ is the radiation flux, which are defined as:
\begin{subequations}
\label{moments}
\begin{equation}
E(0,t)=\frac{1}{2c}\int_{-1}^1I(\mu,0,t)d\mu\equiv a_{\mathrm{Rad}} T_{W}^4(t)
\label{zero_moment}
\end{equation}
\begin{equation}
F(0,t)=\frac{1}{2}\int_{-1}^1I(\mu,0,t)\mu d\mu\equiv \dot{E}_W(t)
\end{equation}
\end{subequations}
$c$ is the speed of light and $a_{\mathrm{Rad}}$ is the radiation constant ($a_{\mathrm{Rad}}=4\sigma_{\mathrm{sb}}/c$). We note that $E_W(t)$ was derived explicitly by the self-similar solution - Eq.~\ref{energy}. Substituting Eq.~\ref{intensity} in Eq.~\ref{inc_F1}, using the definitions of Eqs.~\ref{moments} yields:
\begin{equation}
F_{\mathrm{inc}}(0,t)=\frac{c}{4}E(0,t)+\frac{1}{2}F(0,t)  .
\end{equation}
Using the definitions of $E(0,t)$ and $F(0,t)$, Eqs.~\ref{moments}, yields the relation between $T_{\mathrm{Rad}}$ and $T_W$ (with the help of the self-similar Eq.~\ref{energy}):
\begin{equation}
\sigma_{\mathrm{sb}} T_{\mathrm{Rad}}^4(t)=\sigma_{\mathrm{sb}} T_{W}^4(t)+\dot{E}_W(t)/2
\label{2temp}
\end{equation}

\subsection{Final equations}
\label{fin_sec}

We summarize briefly the final procedure for yielding $T_{\mathrm{Rad}}(D_s)$:
\begin{itemize}
       \item First, we use Eq.~\ref{final_shock} to find the relation $T_0(D_s)$, using the opacity factor calibration Eq.~\ref{calib} and the Hugoniot relations of Al Eq.~\ref{r2_final}. We use the coefficients $p_0(\uptau)$ and $u_0(\uptau_s(\uptau),r_2(D_S))$ directly from Figs.~\ref{ss_coef}(a) and~\ref{ss_coef2}(a).
       \item Second, we use Eq.~\ref{temperature_pwrlaw} to find $T_W(D_s)$.
       \item Finally, we use Eq.~\ref{2temp} with Eq.~\ref{energy} to find $T_{\mathrm{Rad}}(D_s)$. We use Fig.~\ref{ss_coef}(b) for the constant $e_0(\uptau)$.
\end{itemize}
The final equations are:
\begin{subequations}
\label{final0}
\begin{align}
& T_0(t_{\mathrm{Pulse}})=\left[\frac{2}{r_2(D_S)+2}\cdot\frac{1}{u_0(\uptau_S(\uptau),r_2(D_S))\sqrt{p_0(\uptau)\kappa_0^{P_{\omega_1}}(t_{\mathrm{Pulse}})}}D_S\right] ^{\frac{2}{P_{\omega_2}}}t^{\frac{-\uptau_S(\uptau)}{P_{\omega_2}}}_{\mathrm{Pulse}} \label{final1} \\
& T_W(t_{\mathrm{Pulse}})=T_0(t_{\mathrm{Pulse}})t^{\uptau}_{\mathrm{Pulse}} \label{final2} \\
& T_{\mathrm{Rad}}(t_{\mathrm{Pulse}})=\left[\left(\sigma_{\mathrm{sb}} T_{W}^4(t_{\mathrm{Pulse}})+\frac{e_0(\uptau)E_{\omega_3}(\uptau)}{2}\kappa_0^{E_{\omega_1}}(t_{\mathrm{Pulse}})T_0^{E_{\omega_2}}t^{E_{\omega_3}(\uptau)-1}_{\mathrm{Pulse}}\right)\bigg/\sigma_{\mathrm{sb}}\right]^{\nicefrac{1}{4}} \label{final3}
\end{align}
\end{subequations}
\begin{figure}
\centering{
(a)
\includegraphics*[width=7.5cm]{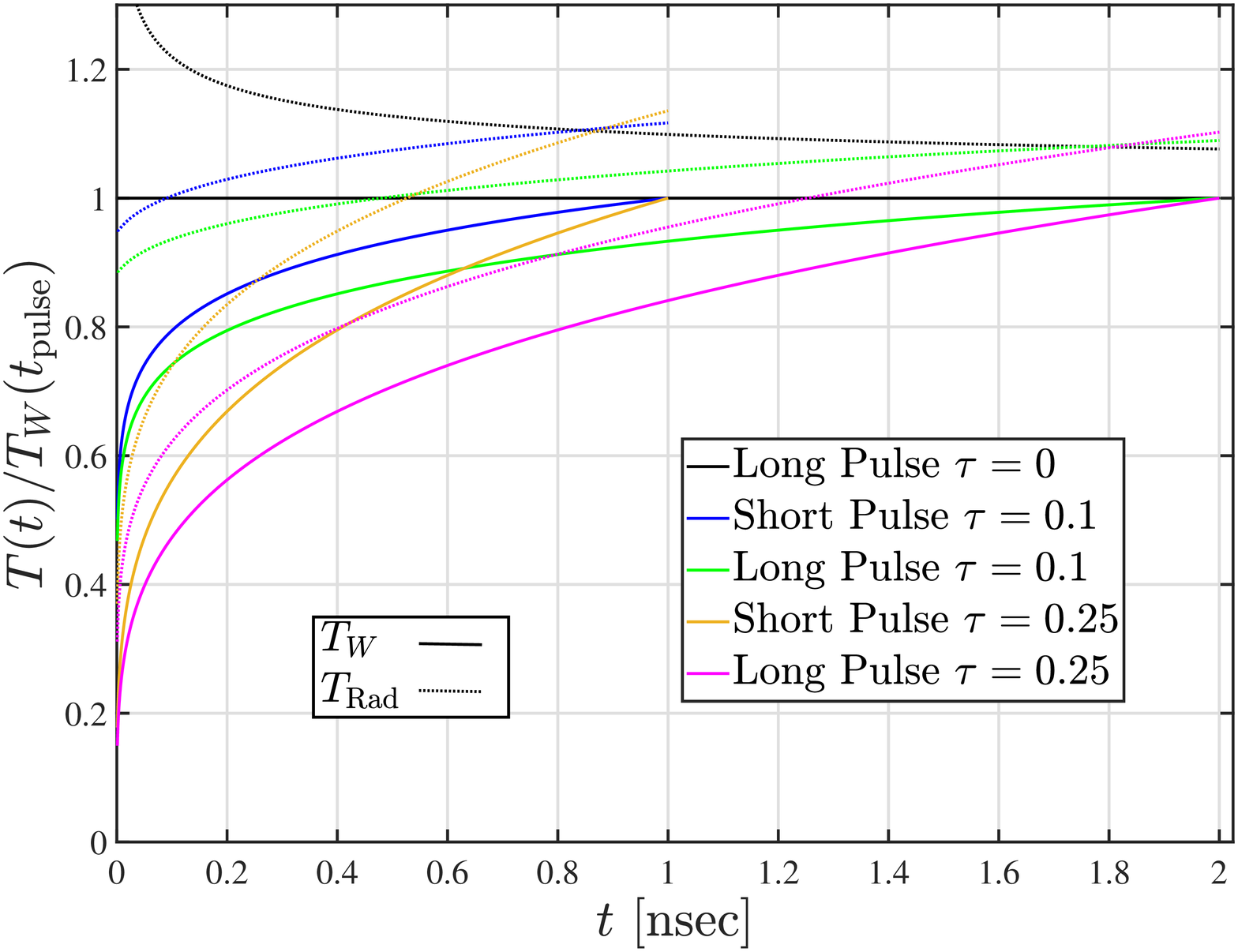}
(b)
\includegraphics*[width=7.5cm]{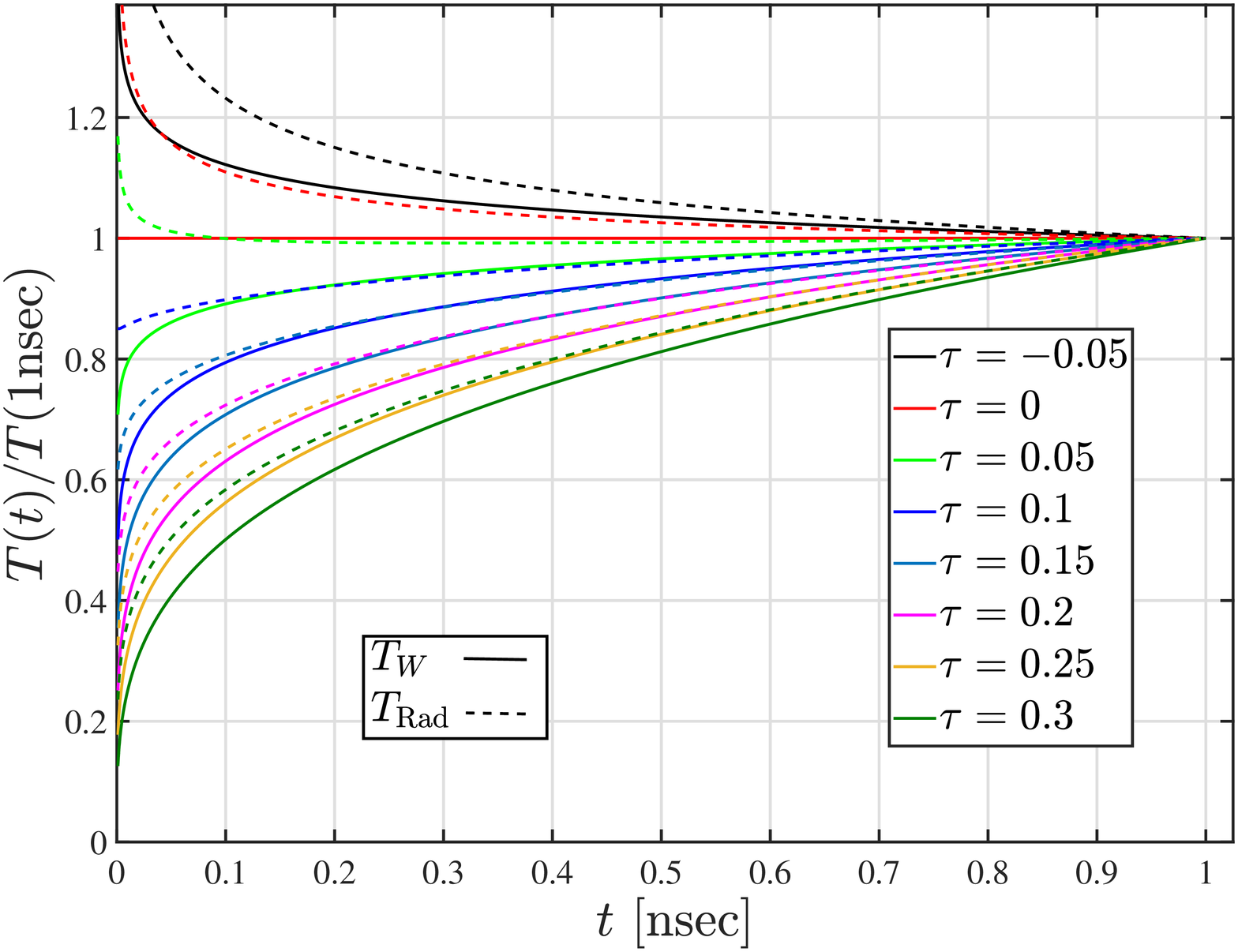}
}
\caption{(Color online) (a) Typical temporal temperature profiles in different duration that were studied theoretically in this work, normalized to the wall temperature $T_W$ in end of the pulse $t_{\mathrm{Pulse}}$. The wall temperature are in solid curves, while we added (dotted curves) for each specified $T_W$ the matching radiation drive temperature $T_{\mathrm{Rad}}$ that represents the incident flux to the wall by Eq.~\ref{final3} (in a typical wall temperature of 200eV). (b) Normalized temporal temperature profiles, wall temperature $T_W$ in solid curves and drive temperatures $T_{\mathrm{Rad}}$ in dashed to $T(t_{\mathrm{Pulse}}=1$nsec). $T_{\mathrm{Rad}}$ temporal behavior matches approximately to $T_W$ with $\uptau_{T_{\mathrm{Rad}}}\approx\uptau-0.05$.}
\label{pulses_the}
\end{figure}

In Fig.~\ref{pulses_the}(a) we plot different typical temperature profiles that we have studied theoretically in this work. The pulses have different time duration (long and short pulses) and different temporal shapes, via $\uptau$, and normalized to the wall temperature $T_W$ at the end of the pulse $t_{\mathrm{Pulse}}$. We can see that as $\uptau$ decreases, the profiles rises more rapidly. The solid curves represent the profiles of the wall temperatures $T_W$, while we add the matching drive temperature $T_{\mathrm{Rad}}$ (the temperature of the incident flux to the wall) for each pulse in dotted curves. The relation between $T_W$ and $T_{\mathrm{Rad}}$ in Fig.~\ref{pulses_the}(a) is via Eq.~\ref{final3} (and~\ref{energy}), using the wall energy in a representative wall temperature of 200eV (thus, the drive temperature is higher at the end of the pulse). We note that the temporal profile of $T_{\mathrm{Rad}}$ slightly differs from the temporal profile of $T_W$. In Fig.~\ref{pulses_the}(b), we plot normalized temperature profiles of both $T_W$ and $T_{\mathrm{Rad}}$ (each one is normalized to its own value at the end of the pulse) in intervals of 0.05. We can see that the temporal behavior of $T_{\mathrm{Rad}}$ nicely matches approximately to $\uptau_{T_{\mathrm{Rad}}}\approx\uptau-0.05$ for all $\uptau$.

\section{Results}
\label{results}

\begin{figure}
\centering{
\includegraphics*[width=15cm]{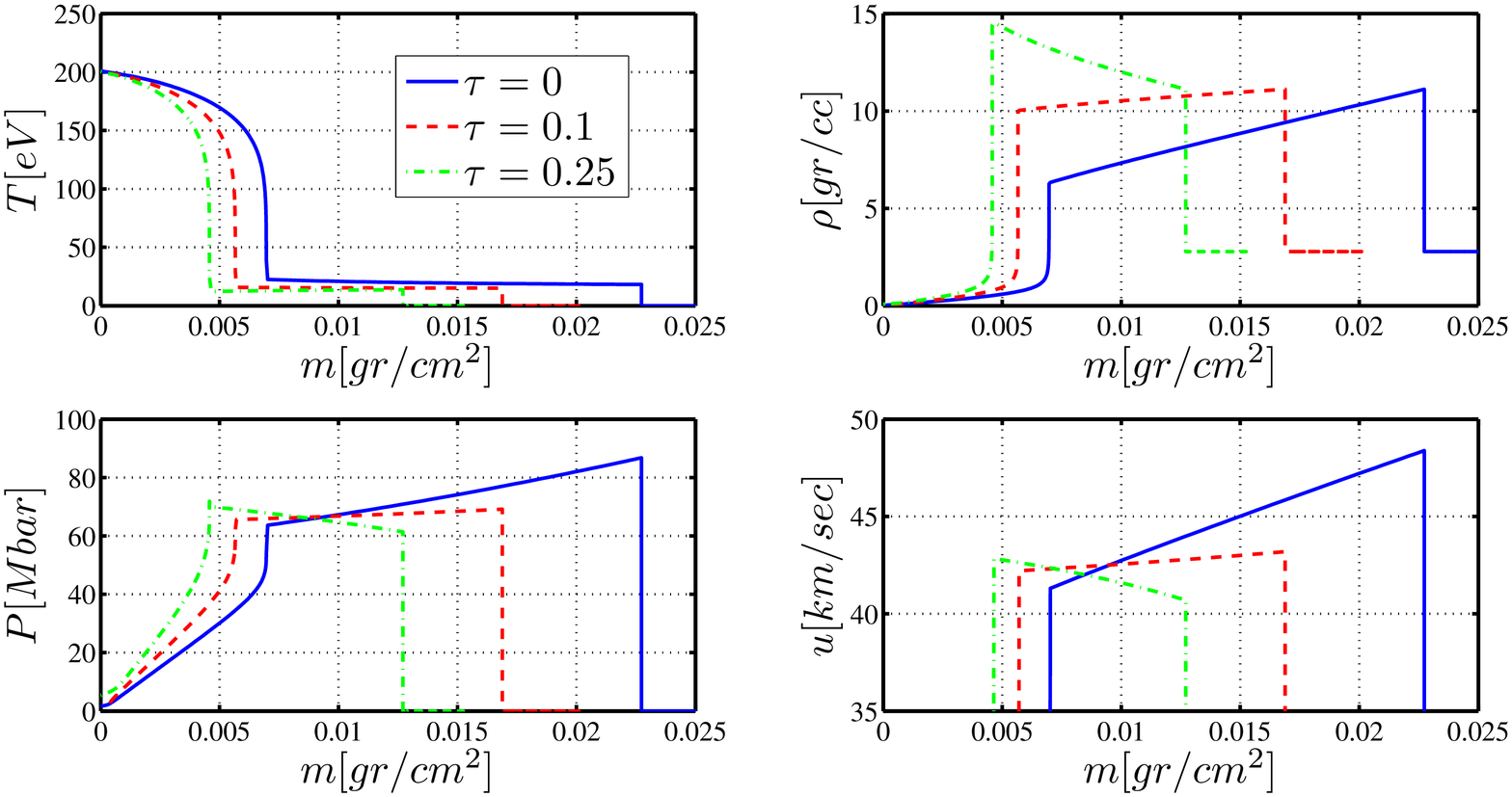}
}
\caption{(Color online) The hydrodynamic profiles of the temperature, the density, the pressure and the velocity as a function of the Lagrangian
coordinate, for different boundary conditions - $\uptau$, using the binary EOS model (The velocity profile is zoomed on the shock region). We used
$r_1=0.3$ (table \ref{table:pwr_law_opac_eos}) and $\kappa_0=1.5$ (typical value for $t_{\mathrm{Pulse}}=1$nsec) for the heat wave region,
and $r_2=0.67$ for the shock region (typical value of Eq. \ref{r2_final}).}
\label{profiles}
\end{figure}

First, we plot (Fig.~\ref{profiles}) the hydrodynamic profiles for a typical example case, using a surface temperature of $T_0=200$eV at 1nsec, using different temporal profiles, $\uptau=0$, $0.1$ and 0.25. We use the binary EOS, with different EOS for the heat and shock regions~\cite{ts2}. For the ablative heat region we use the parameters of Table~\ref{alum_params} and $\kappa_0=1.5$ (for $t=1$nsec), and for the shock region we take $r_2=2/3$ which corresponds to $D_S=50$km/sec (see Fig.~\ref{eos_plot}(b)).

The different regions can be seen clearly, as the heat wave region creates sharp temperature profile, and the material ablates rapidly. This creates strong ablation pressures ($\approx50-80$Mbar), driving a rapid shock wave, which propagates ahead of the heat front. 
We can see that different $\uptau$ yields different shock velocities, with the same maximal $T_W$. Lower $\uptau$ yields a faster heat wave, a stronger pressure profile, and faster particle (and shock) velocities, due to more net energy stored in the sample. Our model therefore, predicts a sensitivity to the temporal profile. We will examine this later versus the experimental results.

\begin{figure}
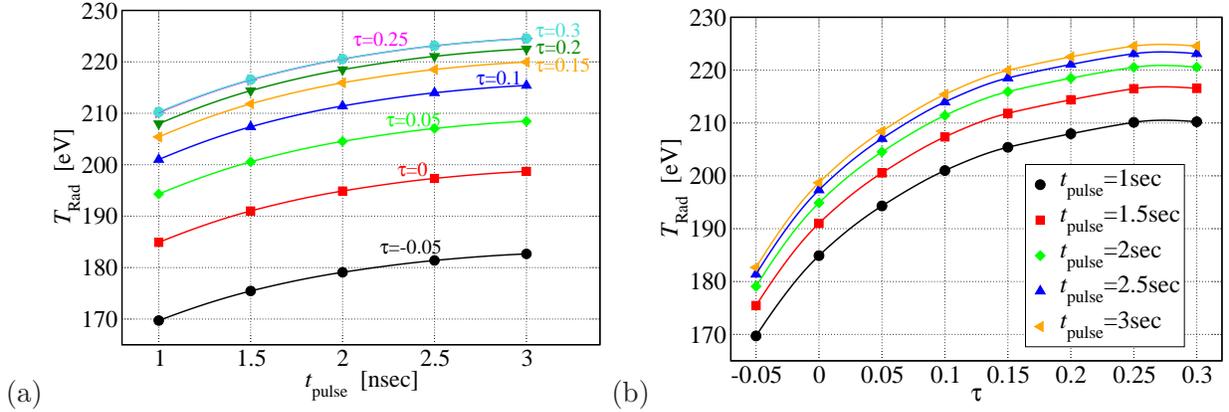

\centering{
(a)
\includegraphics*[width=7.3cm]{t_t.eps}
(b)
\includegraphics*[width=7.5cm]{t_tau.eps}
}
\caption{(Color online) (a) The drive temperature as a function of temperature profile's duration $t_{\mathrm{pulse}}$ for various temporal shapes ($\uptau$). (b) The drive temperature as a function of the temporal shape ($\uptau$) for various values of temperature profile's duration $t_{\mathrm{pulse}}$. In both figures we can see that the curves saturates for $\uptau>0.15$ and $t_{\mathrm{pulse}}>2$nsec.}
\label{results1}
\end{figure}
Using Eqs.~\ref{final0}, we demonstrate in Fig.~\ref{results1} the dependency of both the temperature pulse's duration $t_{\mathrm{pulse}}$ and the temporal profile, via $\uptau$, for a given typical out-going shock velocity of 50km/sec. The important result of Fig.~\ref{results1} is that there is clearly a gap of $\approx10$eV as a function of the temperature pulse's duration $t_{\mathrm{pulse}}$ (exactly the 10eV that Li et al. claim to be the difference between Kauffman's and their results), and $\approx40$eV as a function of the temporal shape, {\bf for a given shock velocity of 50km/sec}. This deviation is of course due to the different amount of energy that is stored inside the sample for different $t_{\mathrm{pulse}}$ or $\uptau$. This demonstrates the strength of our model, which leads to a {\em non} universal $T_{\mathrm{Rad}}(D_S)$ relation, supporting the different fits to the experiments reviewed in Sec.~\ref{eperiments}. Nevertheless, we can see that the $T_{\mathrm{Rad}}$ deviation saturates for $\uptau>0.15$ and $t_{\mathrm{pulse}}>2$nsec.

\begin{figure}
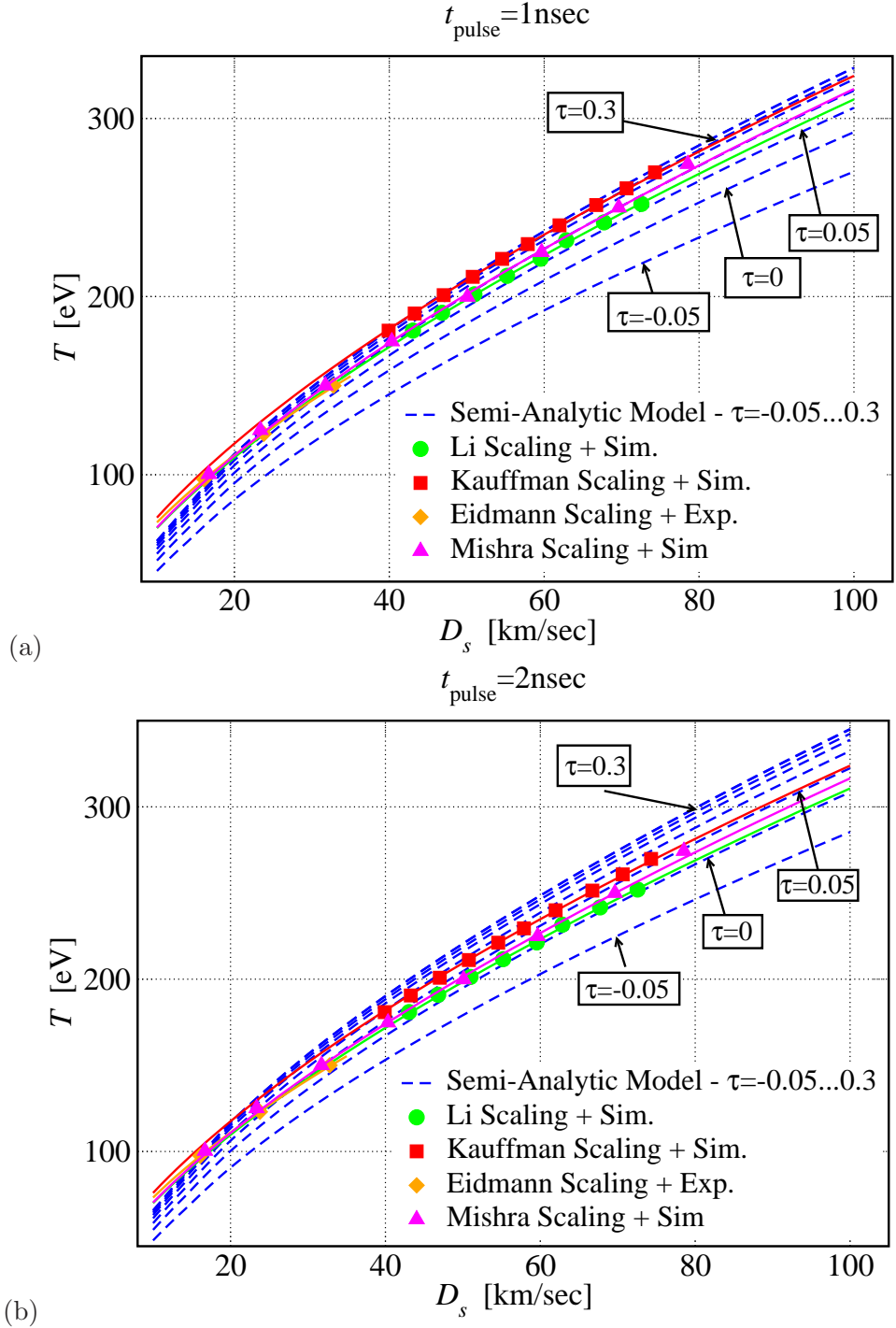

\centering{
(a)
\includegraphics*[width=12cm]{T_D_1nsec.eps}
\\(b)
\includegraphics*[width=12cm]{T_D_2nsec.eps}
}
\caption{(Color online) (a) The radiation temperature $T_{\mathrm{Rad}}$ as a function of the shock velocity $D_S$ for short pulse $t_{\mathrm{pulse}}=1$nsec using the simple semi-analytic model for different values of $-0.05\leqslant\uptau\leqslant 0.3$ (blue curves). For comparison, we have added the different scaling laws from the literature: Kauffman's scaling relation (in red), Li's (green), Eidmann's (orange) and Mishra's (magenta). (b) The same for long pulse, $t_{\mathrm{pulse}}=2$nsec.}
\label{results2}
\end{figure}

In Fig.~\ref{results2} we show the main results using our semi-analytic model. In the dashed blue curves we plot $T_{\mathrm{Rad}}(D_S)$ for different values of $-0.05\leqslant\uptau\leqslant 0.3$, for short pulse (Fig.~\ref{results2}(a)) and long pulse (Fig.~\ref{results2}(b)) experiments. We have added Kauffman's scaling relation (in red), as well as Li's (green), Eidmann's (orange) and Mishra's (magenta).
First, we can see that for a given $D_S$ there is a range of possible $T_{\mathrm{Rad}}$ with a spread of few dozens of eVs, as a function of $\uptau$, similarly to Fig.~\ref{results1}. In addition, longer pulses yield higher $T_{\mathrm{Rad}}$ than short pulses, for a given $D_S$ and $\uptau$. Moreover, the difference between short pulses (Fig.~\ref{results2}(a)) and long pulses (Fig.~\ref{results2}(b)) is $\approx10$eV, again exactly as Li et al. claim to be the difference between Kauffman's and their own results. However, it can be seen that for $\uptau>0.1$, the spread decreases dramatically, and $T_{\mathrm{Rad}}(D_S)$ becomes ``universal", and saturates with $\uptau$.

Second, Fig.~\ref{results2}(a) (short pulse) shows that Kauffman's scaling law is at the {\em upper} limit of the possible blue curves fan, whereas in Fig.~\ref{results2}(b) (long pulse) it is right in the middle of the blue curves, matching $\tau\approx 0.05-0.1$. This is due to the fact that Kauffman's scaling have used {\em long} pulses for calibrating the scaling law~\cite{kauff1,kauff2}, as Li et al. pointed out~\cite{sini1}. Also, vise versa, in Fig.~\ref{results2}(b) (long pulse) Li's scaling law lies at the {\em lower} part of the possible blue curves fan, whereas in Fig.~\ref{results2}(a) (short pulse) it is right in the middle of the blue curves, matching $\tau\approx 0.1-0.15$. This is due to the fact that Li's scaling have used {\em short} pulses for calibrating the scaling law~\cite{sini1}. As mentioned in Sec.~\ref{fin_sec}, $\uptau_{T_{\mathrm{Rad}}}\approx\uptau-0.05$ (see Fig.~\ref{pulses_the}(b)).
Therefore Kauffman's scaling matches to $\uptau_{T_{\mathrm{Rad}}}\approx0-0.05$, and Li's for $\uptau_{T_{\mathrm{Rad}}}\approx0.05-0.1$, This matches the temporal profiles in the experiments quite well (see Fig.~\ref{exp_pul}). The simple semi-analytic model reproduces and explains the difference between Kauffman's and Li's scaling laws that was pointed out in~\cite{sini1,sini2} and previously explained only by full simulations.

Eidmann's scaling law (orange curves) that was developed for the lower temperature range of $100<T_{\mathrm{Rad}}<150$eV~\cite{german}, matches Li's scaling law (green curves) and is quite different from Kauffman's (red curves). This is directly due to the short pulses of 0.8nsec that were used in the Gekko-XII experiments. Next, Mishra's scaling law (magenta curves) is closer to Li's than to Kauffman's~\cite{india}. 
At first sight, this is a contradiction, since Mishra et al. have used long temperature profiles of 3nsec. However, Mishra et al. have used a constant surface temperature as BC, i.e. $\uptau=0$, and this is the reason that their scaling is lower than Kauffman's. Not surprisingly, Mishra's results (magenta) matches almost perfectly our model using $\uptau=0$, in Fig.~\ref{results2}(b) (long pulses).

We also note that Nova's ``two-steps" temperature profile (see black/red curves in Fig.~\ref{exp_pul}), creates a `kink' in the distance traversed by the shock trajectory curve, yielding two different shock-velocities~\cite{remington,india_old,sini1,Das}. The simple analytic model reproduces this fact completely, separating the temperature profile to two pulses with different temperatures. Ghosh et al. presents some quantitative data~\cite{india_old}, showing that the shock velocity changes from 35.4 to 54.6km/sec, and the first step is characterized by an incident radiation temperature of $T_{\mathrm{Rad}}\approx140$eV, which increases to 200eV after 1.5nsec. Fig.~\ref{results2} reproduces these results, using $\uptau\approx0-0.05$ (this experimental result is less sensitive to the temperature profile's duration).

\begin{figure}
\centering{
\includegraphics*[width=8.5cm]{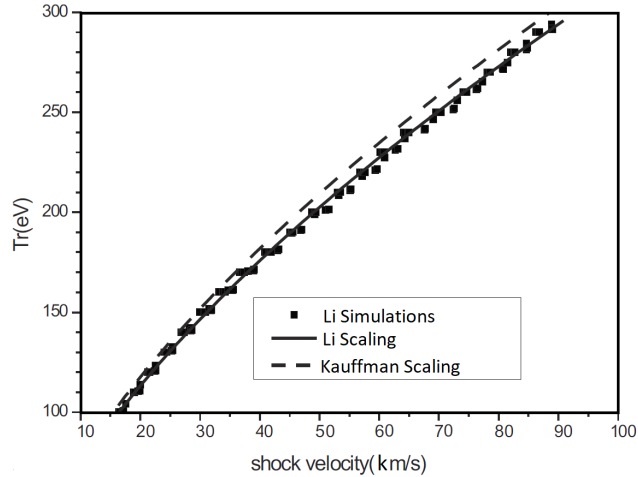}
}
\caption{(Color online) The drive temperature $T_{\mathrm{Rad}}$ as a function of shock velocity in Li et al. simulations using SG-II and SG-III pulses. Kauffman's (dashed) and Li's (solid) scaling relations are presented as well. Reproduced from~\cite{sini1}, with the permission of AIP Publishing.}
\label{results_sini}
\end{figure}

The simple model predicts that the drive temperature $T_{\mathrm{Rad}}(D_S)$ curve is a function of both the duration and the temporal profile of the drive temperature pulse. However, Li et al. claim that the $T_{\mathrm{Rad}}(D_S)$ curve is duration-dependent, with a difference of $\approx10$eV between short (1nsec) and long (2-2.5nsec) pulses, but the influence of the {\bf temporal profiles} on the scaling relation is
negligible, for x-ray sources driven by 1nsec laser. This apparent contradiction may be solved easily. The temporal profile of the SG-III experiment may be approximated by $\uptau\approx 0.1-0.15$ and that of SG-II by $\uptau\approx0.35$ (see Fig.~\ref{exp_pul} and Sec.~\ref{eperiments}). In this regime of $\uptau>0.15$, $T_{\mathrm{Rad}}(D_S)$ is indeed almost universal, and saturates, as can be seen in both Figs.~\ref{results1} and~\ref{results2}. Moreover, in Fig.~\ref{results_sini} which is taken from~\cite{sini1}, the difference of $\approx10$eV between the two scaling laws, which represent the different profiles durations, may be seen easily. But, we can see the simulations scatter about $\approx5$eV around the temporal profile. This is exactly the $\approx5$eV difference that can be seen in Figs.~\ref{results1} and~\ref{results2} for $\uptau>0.15$. I.e.,  expanding Li et al. research to lower $\uptau$s or different temporal shapes, would lead to a major discrepancy due to the different temporal shapes.

\section{Conclusion}
\label{discussion}

Ablative subsonic radiative heat waves, or the subsonic Marshak waves have been studied for three decades. The coupling between the hydrodynamic equations and the radiative heat transfer is crucial for modeling hohlraums, in the quest for indirect drive ICF~\cite{lindl2004,rosen1996}. This feature enables us to evaluate the radiation drive temperature $T_{\mathrm{Rad}}$ inside the hohlraum, through the measurement of the velocity of the emitted shock wave $D_S$ that is developed in a well-characterized material, such as aluminium~\cite{kauff1,kauff2,hatchett,campbell,german,sini1}.

Recently, we have derived a basic theoretical study, yielding a full self-similar solution for the subsonic problem, patching two self-similar solutions, one for the heat region and one for the shock region~\cite{ts1}. We have expanded this basic model to include more complex material behavior, i.e., EOS, since the EOS properties of the shock and the heat regions, are very different~\cite{ts2}. 

This study takes the basic theoretical study of ablative subsonic radiative waves, and confront it versus the results of the various experiments in which the actual relation between the heat region (via $T_{\mathrm{Rad}}$) and the shock region (via $D_S$) was measured. The different studies usually modeled this problem with direct radiative-hydrodynamics simulations, yielding different scaling relations between $T_{\mathrm{Rad}}$ and $D_S$~\cite{kauff1,sini1,german,india} (see Sec.~\ref{eperiments}). The current study enables us to test this issue semi-analytically (via the self-similar solutions), and is aimed to understand the differences between the different scaling laws.

The model is derived in details for aluminium in Sec.~\ref{model}, taking into account the transport effects via calibration of an opacity factor by IMC simulations in the heat region, and the detailed Hugoniot relations~\cite{hugo1}, in the shock region. We have found that the $T_{\mathrm{Rad}}(D_s)$ relation {\em is not} universal, and depends on the different features of the temporal behavior of the temperature pulse, such as the duration $t_{\mathrm{pulse}}$, and its temporal profile (through $\uptau$), with up to $\approx 40$eV diversity in $T_{\mathrm{Rad}}$. This diversity is due to the different total energy that is stored inside the aluminium sample for the different conditions of the profile's parameters. This explains the difference between the different scaling laws found in the literature. Moreover, the simple model recovers the different experimental and simulation data, separating short (Fig.~\ref{results2}(a)) and long (Fig.~\ref{results2}(b)) pulses, each for different temporal profiles ($\uptau$). Specifically the model explains the difference between Kauffman's and Li's scaling laws, by simple analytic expressions.

The simple model enables an estimate of $T_{\mathrm{Rad}}(D_s)$ for any future experiments using a power-law fit of the drive temperature temporal profile and the temperature pulse's duration, using the expressions of Eqs.~\ref{final0}, for aluminium, with which the vast majority of experiments were performed. We plan to expand the research in the future for different materials, depending on the advancement of the research and experimental program in this field.

\end{document}